# Homogenization and Clustering as a Non-Statistical Methodology to Assess Multi-Parametrical Chain Problems

Johannes Freiesleben and Nicolas Guérin

**Abstract**

We present a new theoretical and numerical assessment methodology for a one-dimensional process chain with general applicability to management problems such as the optimization of decision chains or production chains. The process is thereby seen as a chain of subsequently arranged units with random parameters influencing the objective function. For solving such complex chain problems, analytical methods usually fail and statistical methods only provide approximate solutions while requiring massive computing power. We took insights from physics to develop a new methodology based on homogenization and clustering. The core idea is to replace the complex real chain with a virtual chain that homogenizes the involved parameters and clusters the working units into global units to facilitate computation. This methodology drastically reduces computing time, allows for the derivation of analytical formulas, and provides fast and objective insights about the optimization problem under investigation. We illustrate the analytical potency of this methodology by applying it to the production problem of selecting the economically superior quality maintenance strategy. It can further be applied to all sequential multi-parametrical chain problems commonly found in business.

## 1. Introduction

Business reality is characterized by processes. This is most obvious in the production division of a company with its focus on generating standardized products or services, but no less relevant for the administrative part and managerial systems, which in most of its routine tasks also follow the logic of repeat processes. It has been stated that changing environments require an adaptive organization where over-standardization is counterproductive and might inhibit necessary change (e.g. Benner and Tushman, 2003), but it is still evident that a certain degree of standardized processes will even survive in a high customization environment simply for cost reasons (e.g. Duray et al, 2000). With most business processes being of technical nature and repeat-based by design, they are mostly sequential processes, meaning a chain of sequentially ordered process steps, each one being dependent on the output of the former in its value creation. Highly complex process landscapes can be split into shorter and more simple sequential processes to ease manageability and analysis.

Analysis of even such simple chains, however, is often complicated due to the complexity of the involved decision parameters. Chain problems are seldom one-parametrical, and even if they are the parameter might manifest itself in a power law form, rendering simple analytical derivation of solutions impossible. The common way of finding solutions to these complex power law functions are statistical modeling or approximation methods, such as Monte Carlo Simulations or Genetical Algorithms, to name just two prominent examples (e.g. Aytug et al, 2003; Harrison et al, 2007). Involving statistical ranges based on expected mean values and probability distributions gives acceptable solutions to practical problems, as has been shown time and again, but they require long computing time (e.g. Gentle, 2012) and yield no direct results, but rather approximate solutions that in their quality depend on the skillful modeler to set the correct initial settings (e.g. Law, 2015).

We want to present in this paper a novel methodology to solve such complex chain problems, which works without statistical values and can be conducted with minimal computing time. It is based on the idea of homogenization lending originally from the principle of canonical transformation in Physics (e.g. Landau and Lifshitz, 1975). The homogenization technique is widely practiced in material science to model complex composite materials in mechanics (e.g. Takano, 2000) as well as in electromagnetism (e.g. Alù, 2011). Its objective could be outlined as "... to find some kind of equivalent material model without needing to represent each individual microstructure. This model should characterize the average behavior as well as represent the effect of the composite material heterogeneities..." (Guedes, 1990, pp 143). Its purpose thus is to reduce the complexity in a multi-parameter setting by assuming a uniform distribution of parameter values. We will show that this methodology can be universally applied to multi-parametrical chain problems and thus facilitate their solution.

In order to facilitate the computation with the now homogenized parameter values, we develop different clustering scenarios of the chain regarding the chain length itself as a variable. We assess these clustering scenarios and develop theorems on their applicability. To test our new approach, we apply it to a practical chain problem common to the production array - the selection of the economically superior quality maintenance strategy - and compute both theoretical results derived from our approach and numerical results from a given example set of parameter values. We find that the results are of high value for the problem solution as they provide relevant insights into the optimization issue. To our knowledge, there is no methodology available so far that produces equally reliable results while requiring such short computing time.

We take the selection of the optimal quality maintenance strategy, with defect propensities and maintenance effectiveness as the main parameters, as our illustrative case. For this, we develop production cost functions for zero maintenance (baseline strategy), inspection (traditional strategy) and in-process monitoring (technologically advanced strategy) and compare their results. We take this case example as maintenance strategies play an important economic role in virtually any business and are dependent on a variety of parameters, making them a great practical illustration for our results.

The paper is organized as follows. In section 2, the production cost model for the comparison of quality maintenance strategies is developed and the illustrative case introduced. In section 3, we then introduce the concepts of homogenization and clustering and develop our methodological approach. This new methodology is then applied to our illustrative case and discussed in detail in sections 4 and 5, each section covering a different strategy comparison and highlighting the explanatory power of the methodology. The concluding section 6 combines the results from the previous two sections, discusses the significance of our methodology and provides an outlook on further study.

## 2. Quality Maintenance Production Costs

Quality maintenance plays a vital part in production planning. Paraphrasing Juran (1989), quality maintenance comprises "all efforts conducted to safeguard the current level of production quality during the course of production", in other words its objective is to keep the defect rate stable and - ideally - at a low level. As quality maintenance does not come for free, an economic comparison of its technological alternatives seems interesting from an economic point of view. Whereas quality improvement has been the focus of many economic studies (e.g. Zu *et al*, 2008; Chao *et al*, 2009), quality maintenance has received relatively little attention aside from the economic discussion of inspection in the operations research literature (e.g. Raz *et al*, 2000). This is insofar surprising as advances in sensor technology today provide companies with at least one more technological alternative to inspection, which is process monitoring (e.g. Fortuna *et al*, 2005). Papers discussing economic aspects of quality maintenance include empirical studies focused on its strategic importance (e.g. Madu, 2000), its managerial aspects (e.g. Muchiri *et al*, 2010), or its effect on financial performance (e.g. Alsyouf, 2007), but an economic comparison of its main technological options has so far not been put forth. This is why we chose this topic as our illustrative case.

In deciding on quality maintenance, a company can follow one of three principal strategies: inspecting the output of process stages, monitoring quality-relevant process parameters (QRPs) inside the stages, or not investing in maintenance at all (zero maintenance).

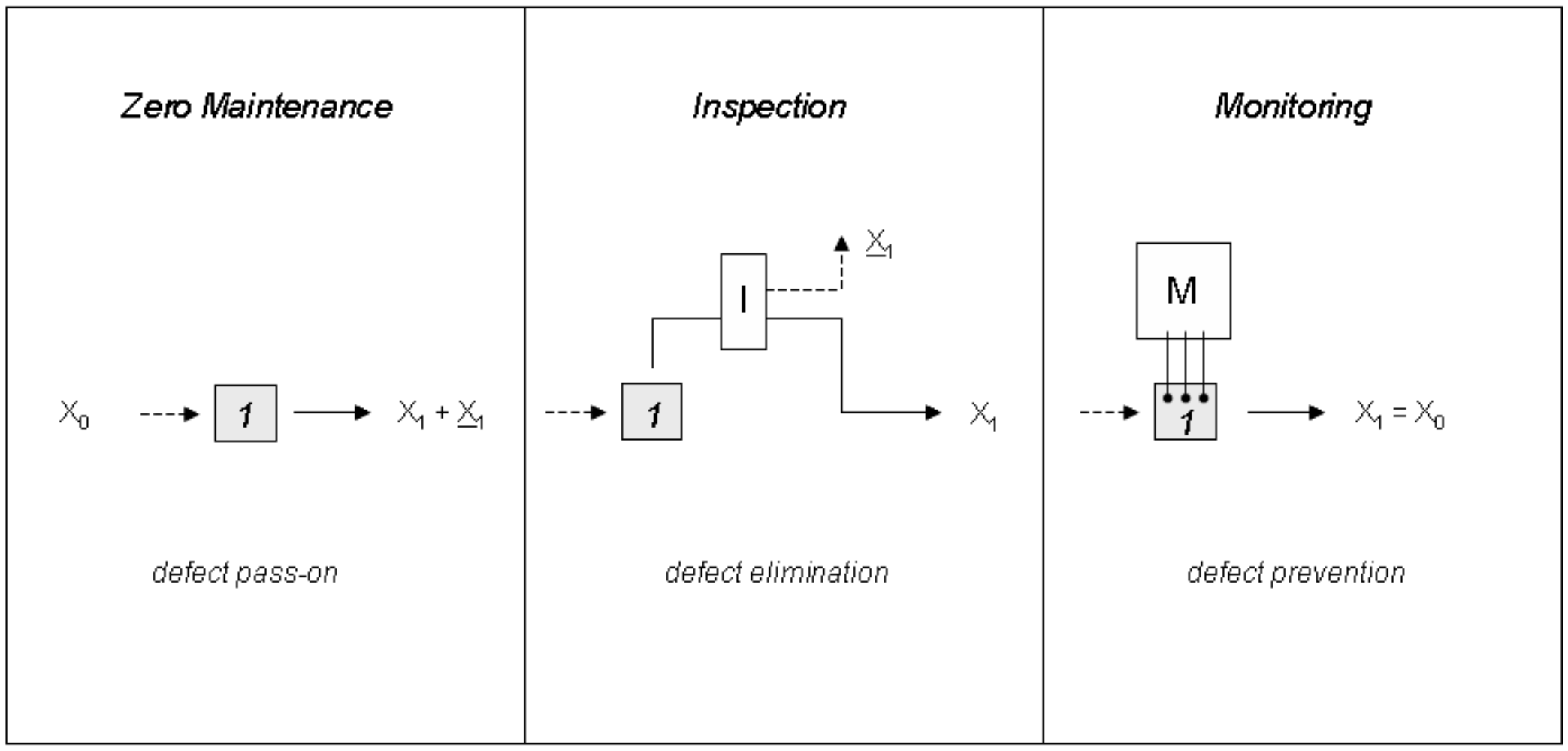

Figure 1: Schematic depiction of zero maintenance, inspection and monitoring.

As figure 1 depicts schematically, all three strategies have different effects on processed volumes. The initial input volume $X_0$ to production stage 1 is partially made defective, but zero maintenance would pass all defective units ($\underline{X}_1$) as well as the good units ($X_1$) undifferentiated on to the next stage, where they are further processed. Inspection in stage 1 would eliminate the defective units ($\underline{X}_1$) such that only good units ($X_1$) are passed on to the next stage. If the production is subjected to monitoring, defects would be prevented by close observation of QRPs and timely counteractions, such that the processed volume $X_0$ is entirely transformed into good output $X_1$. Not included in this schematic illustration is the potential ineffectiveness of both inspection and monitoring, which might result in defective units being declared as “good” and passed-on to the next stage.

We will base our analyses on a simplified chain model of production, consisting of *n* linearly aligned process stages (based on the general model proposed in Freiesleben, 2005). For notations, see table 1. The output of a process stage is input for the next, such that the output of the last process stage is dependent on the outputs of all former stages. In perfect quality condition, the number $X_0$ of units which are transformed in stage 1 is equal to the number of sellable output $X_n$ which leaves the process after stage *n*. As soon as there is a defect rate $d_k$ in any process stage *k*, only the fraction $(1 – d_k)$ of products passed on to the next process stage is of acceptable quality. Inspection aims at separating the defective units from the production flow, as to prevent that additional transformations are applied to already faulty and unsellable units. Therefore, inspection unambiguously reduces $X_n$. Monitoring, on the other hand, aims at detecting potential causes for defects before actual defects are produced, thus keeping the processed and sold volumes equal. However, both maintenance strategies might vary in their effectiveness: inspection might not detect all defective units ($e_i \leq 1$) and monitoring might not keep track of all relevant QRPs ($e_m \leq 1$), resulting in a fraction $\underline{X_n}$ of the sellable-classified output $X_n$ in truth being defective and causing an external quality effect by reaching the customer. Hence, an important parameter in determining the advantageousness of these strategies is their assumed effectiveness. As it is usually below 1, we can state that for inspection $X_0 > X_n > (X_n - \underline{X_n})$ and for monitoring $X_0 = X_n > (X_n - \underline{X_n})$.

| | |
|---|---|
| $\alpha$ | average rate of return of defective products that reached the customer |
| $\beta$ | relative premium over regular $c_u$ as proxy for transaction and goodwill costs |
| $c$ , $c_k$ | variable production cost (general, per stage *k*) |
| $c_u$ | unit costs after completion of production |
| $c_i$ | homogenized $c$ if inspection strategy is chosen |
| $c_m$ | homogenized $c$ if monitoring strategy is chosen |
| $d$ , $d_k$ | defect rate (general, per stage *k*) |
| $e_i$ , $e_{ik}$ | effectiveness of inspection (general, per stage *k*) |
| $e_m$ , $e_{mk}$ | effectiveness of monitoring (general, per stage *k*) |
| $i$ , $i_k$ | variable inspection cost (general, per stage *k*) |
| $m$ , $m_k$ | variable monitoring cost (general, per stage *k*) |
| $C$ , $C_k$ | fixed production cost (general, per stage *k*) |
| $I$ , $I_k$ | fixed inspection cost (general, per stage *k*) |
| $M$ , $M_k$ | fixed monitoring cost (general, per stage *k*) |
| $n$ | number of production stages |
| $N$ | number of virtual production stages, depending on clustering strategy |
| $X_0$ | initial volume processed in production stage 1 |
| $X_n$ | sellable-classified output after stage *n* |
| $\underline{X_n}$ | fraction of $X_n$ in truth being defective due to $e_i, e_m < 1$ |

Table 1: Notations used in the model.

Based on defect rate and effectiveness, a general volume formula for the sold volume $X_n$ as well as the defective sold volume $\underline{X}_n$ can be proposed as

(1) $$X_n = X_0 \prod_{k=1}^{n} \left(1 - e_{ik}\left(1 - e_{mk}\right) d_k\right)$$

(2) $$\underline{X}_n = X_0 \left[ \prod_{k=1}^{n} \left(1 - e_{ik}\left(1 - e_{mk}\right) d_k\right) - \prod_{k=1}^{n} \left(1 - \left(1 - e_{mk}\right) d_k\right) \right]$$

PROOF. See the appendix.

For the two technological maintenance strategies (inspection indicated by superscript "i" and monitoring by "m") and the strategy of zero maintenance (superscript "z") follow volume formulas directly derived from (1) and (2):

(3) $$X_n^i = X_0 \prod_{k=1}^{n} \left(1 - e_{ik} d_k\right) \qquad \underline{X}_n^i = X_0 \left[ \prod_{k=1}^{n} \left(1 - e_{ik} d_k\right) - \prod_{k=1}^{n} \left(1 - d_k\right) \right]$$

(4) $$X_n^m = X_0 \qquad \underline{X}_n^m = X_0 \left[ 1 - \prod_{k=1}^{n} \left(1 - \left(1 - e_{mk}\right) d_k\right) \right]$$

(5) $$X_n^z = X_0 \qquad \underline{X}_n^z = X_0 \left[ 1 - \prod_{k=1}^{n} \left(1 - d_k\right) \right]$$

These volume formulas are the basis for the economic comparison of the three maintenance strategies. A meaningful economic comparison has to focus on the *unit costs per sold product*, since only sold products generate revenues and can contribute to pay for the incurred total costs of production. Part of the sellable-classified volume might be defective due to ineffective maintenance. A strategy of zero maintenance reduces the direct maintenance costs to nil but allows all defective units to be sold, thus generating a negative reputation effect. This effect is difficult to quantify, but we take the warranty costs of the producer as a proxy for this effect. For calculating the warranty costs, we assume an average rate of return $\alpha$ of defective products (sent back by disappointed customers) and a premium of $\beta$ over regular variable production costs, which includes transaction costs for sending the replacement product as well as goodwill costs (costs incurred to restore the goodwill of the disappointed customer, e.g. free give-aways). Note that the "reputation parameters" $\alpha$ and $\beta$ are independent of the defect rate as $d$ is rarely observable for the individual customer and hence has no influence on his general behavior or goodwill expectation. With $C_{fix}$ being the regular fixed costs of production, $C_{var}$ the incurred sum of all variable costs of production and $C_{wty}$ the total warranty costs, the total costs $C_{tot}$ can be proposed as

(6) $$C_{tot} = C_{fix} + C_{\text{var}} + C_{wty}$$ with

(6a) $$C_{fix} = \sum_{j=1}^{n}\left(C_j + M_j + I_j\right)$$

(6b) $$C_{var} = X_0 \sum_{j=1}^{n}\left(\left(c_j + m_j + i_j\right)\prod_{k=1}^{j-1}\left(1 - e_{ik}\left(1 - e_{mk}\right)d_k\right)\right)$$

(6c) $$C_{wty} = \alpha\left(1+\beta\right)\frac{C_{\text{var}}}{X_n}\underline{X_n}$$ where $\alpha \in [0,1]$ and $\beta \in [0,+\infty[$

Substituting $\Theta_C = \prod_{k=1}^{n}\left(1 - e_{ik}\left(1 - e_{mk}\right)d_k\right)$, the general formula for unit costs $c_u$ becomes

(7) $$c_u \equiv \frac{C_{tot}}{X_n} = \frac{\sum_{j=1}^{n}\left(C_j + M_j + I_j\right)}{X_0 \Theta_C} +$$

$$\left(1 + \alpha\left(1+\beta\right)\frac{\Theta_C - \prod_{k=1}^{n}\left(1 - \left(1 - e_{mk}\right)d_k\right)}{\Theta_C}\right)\frac{\sum_{j=1}^{n}\left(\left(c_j + m_j + i_j\right)\prod_{k=1}^{j-1}\left(1 - e_{ik}\left(1 - e_{mk}\right)d_k\right)\right)}{\Theta_C}$$

By application of eq. (7) to each strategy, the unit costs for the three maintenance strategies - pure inspection, pure monitoring, zero maintenance - follow as:

(8) $$c_u^i = \frac{\sum_{j=1}^{n}\left(C_j + I_j\right)}{X_0 \prod_{k=1}^{n}\left(1 - e_{ik}d_k\right)} +$$

$$\left(1 + \alpha\left(1+\beta\right)\frac{\prod_{k=1}^{n}\left(1 - e_{ik}d_k\right) - \prod_{k=1}^{n}\left(1 - d_k\right)}{\prod_{k=1}^{n}\left(1 - e_{ik}d_k\right)}\right)\frac{\sum_{j=1}^{n}\left(\left(c_j + i_j\right)\prod_{k=1}^{j-1}\left(1 - e_{ik}d_k\right)\right)}{\prod_{k=1}^{n}\left(1 - e_{ik}d_k\right)}$$

(9) $$c_u^m = \frac{1}{X_0}\sum_{j=1}^{n}\left(C_j + M_j\right) + \left(1+\alpha\left(1+\beta\right)\left(1-\prod_{k=1}^{n}\left(1-\left(1-e_{mk}\right)d_k\right)\right)\right)\sum_{j=1}^{n}\left(c_j + m_j\right)$$

(10) $$c_u^z = \frac{1}{X_0}\sum_{j=1}^{n}C_j + \left(1+\alpha\left(1+\beta\right)\left(1-\prod_{k=1}^{n}\left(1-d_k\right)\right)\right)\sum_{j=1}^{n}c_j$$

## 3. Homogenization and Clustering as a New Methodological Approach

Practitioners knowing all parameters of eq. (8) to (10) can compute and compare the unit costs resulting from the three maintenance strategies as to decide which one is the most economical. Eq. (8) and (9) contain 6*n*+1 free parameters and eq. (10) contains 3*n*+1 free parameters, which potentially lead to millions of simulations if we consider an extensive numerical study scanning a large range of values for each free parameter. Additionally, such highly heterogeneous formulas (8) to (10) do not allow a direct analytical approach from which general results could be derived. To circumvent these problems, one possible strategy is to define virtual homogeneous chains that replace eq. (8) to (10) and virtual homogeneous parameters $C$, $I$, $M$, $d$, $e_i$, $e_m$, $c$, $i$ and $m$ that replace the free parameters $C_k$, $I_k$, $M_k$, $d_k$, $e_{ik}$, $e_{mk}$, $c_k$, $i_k$ and $m_k$. For example, all $d_k$ taking different values will be replaced by only one parameter $d$ which keeps the same homogeneous value whatever the position *k* in the chain. Thus, *n* values are compacted into one value that can be regarded as a kind of "mean value" of all $d_k$. The required condition is that all homogeneous parameters keep the final results unchanged, i.e. the final volumes $X_n$ and $\underline{X}_n$ and the total cost values resulting from eq. (6a) to (6c) remain unchanged. The advantage of such an homogenization technique is that the cost functions (8) and (9) will be replaced by homogeneous cost functions having only 7 free parameters, and (10) will after transformation have only 4 free parameters. This is a practice common in physics when instable statistical parameters are considered (e.g. Landau and Lifshitz, 1975). Table 2 lists the canonical transformations replacing inhomogeneous parameters by its respective homogeneous form keeping final volumes and costs unchanged.

| | Parameter | Homogenized formulas | Boundary |
|---|---|---|---|
| (11) | Fixed Costs (C,M,I) | $V\left(N\right)=\frac{1}{N}\sum_{j=1}^{n}V_j$ | $V\left(N\right)\geq 0$ |
| (12) | Defect Rate | $d\left(N\right)=1-\left[\prod_{k=1}^{n}\left(1-d_k\right)\right]^{1/N}$ | $d\left(N\right)\in[0,1]$ |
| (13) | Effectiveness Maintenance | $e_m\left(N\right)=1-\frac{1-\left[\prod_{k=1}^{n}\left(1-\left(1-e_{mk}\right)d_k\right)\right]^{1/N}}{1-\left[\prod_{k=1}^{n}\left(1-d_k\right)\right]^{1/N}}$ | $e_m\left(N\right)\in[0,1]$ |
| (14) | Effectiveness inspection | $e_i\left(N\right)=\frac{1-\left[\prod_{k=1}^{n}\left(1-e_{ik}\left(1-e_{mk}\right)d_k\right)\right]^{1/N}}{1-\left[\prod_{k=1}^{n}\left(1-\left(1-e_{mk}\right)d_k\right)\right]^{1/N}}$ | $e_i\left(N\right)\in[0,1]$ |
| (15) | Variable Costs (c,m,i) | $v\left(N\right)=\sum_{j=1}^{n}\left(v_j\prod_{k=1}^{j-1}\left(1-e_{ik}\left(1-e_{mk}\right)d_k\right)\right)\frac{e_i\left(1-e_m\right)d}{1-\left(1-e_i\left(1-e_m\right)d\right)^N}$ | $v\left(N\right)\geq 0$ |

Table 2: Homogenized formulas for equation parameters and related boundaries

Additionally to homogenization, the analytical power can be greatly advanced by "virtualization" of the original production string. Let $N$ be the number of virtual process stages over which to uniformly spread parameter values of a heterogeneous production string with $n$ real stages. The heterogeneous unit costs functions (7) to (10) thus become homogeneous as follows:

(16) $$c_u=\frac{C_{TOT}}{X_n}=N\frac{C+M+I}{X_0\left(1-e_i\left(1-e_m\right)d\right)^N}+$$
$$\left(c+m+i\right)\frac{1-\left(1-e_i\left(1-e_m\right)d\right)^N}{e_i\left(1-e_m\right)d\left(1-e_i\left(1-e_m\right)d\right)^N}\left(1+\alpha\left(1+\beta\right)\frac{\left(\left(1-e_i\left(1-e_m\right)d\right)^N-\left(1-\left(1-e_m\right)d\right)^N\right)}{\left(1-e_i\left(1-e_m\right)d\right)^N}\right)$$

(17) $$c_u^i=N\frac{C+I}{X_0\left(1-e_id\right)^N}+\left(1+\alpha\left(1+\beta\right)\frac{\left(1-e_id\right)^N-\left(1-d\right)^N}{\left(1-e_id\right)^N}\right)\left(c_i+i\right)\frac{1-\left(1-e_id\right)^N}{e_id\left(1-e_id\right)^N}$$

(18) $$c_u^m = N\frac{C+M}{X_0} + N\left(c_m + m\right)\left(1+\alpha\left(1+\beta\right)\left(1-\left(1-\left(1-e_m\right)d\right)^N\right)\right)$$

(19) $$c_u^z = N\frac{C}{X_0} + Nc_m\left(1+\alpha\left(1+\beta\right)\left(1-\left(1-d\right)^N\right)\right)$$

where $c_i$ denotes the homogenized $c$ for inspection [eq. (15) without $e_m$] and $c_m$ the homogenized $c$ for monitoring and zero maintenance [eq. (15) without $e_i$]. Any heterogeneous total cost function with varying parameters per stage can thus be transformed into a strictly homogeneous function by canonical transformation, i.e. by applying (11) to (15) there is no difference in numerical values between unit costs computed from the heterogeneous form [(7) to (10)] and the homogeneous form [(16) to (19)]. This allows any real-life production process to be analyzed by applying the homogenized form.

Of importance is the free rescaling parameter $N$, as the process can be thought of as being "clustered" into one single virtual process stage $N = 1$, which is homogeneous by definition, or spread over any number of homogenized stages. There are thus three general scaling approaches of $N$:

1) $N < n$ clustering of the chain (smaller virtual than real chain)
2) $N = n$ direct length representation (virtual is equal to real chain)
3) $N > n$ de-clustering of the chain (larger virtual than real chain)

$N$ thus offers an additional degree of freedom for analyzing the chain, with different implications to be discussed in the next section. An economic comparison of maintenance strategies means determining which of eq. (16) to (19) yields the lowest unit costs given the involved parameters. For this aim, we need to transform these non-linear into more simple equations, which can be achieved by the outlined homogenization and clustering methodology.

Regarding first the choice of the free parameter $N$, the different scaling approaches have different methodological implications. Direct length representation ($N = n$) rarely yields analytical solutions due to the involved power laws. By additionally applying Taylor approximation at the first order this problem might be alleviated but the conditions for such an approximation have to be set properly (for instance, for comparisons involving inspection or zero maintenance, any high defect rate will make a Taylor treatment impossible). De-clustering ($N > n$) may look favorable *a priori* as the homogenized parameters get smaller with $N$, so that the Taylor approximation may apply more easily, but as the power index $N$ gets bigger, the final error potentially gets bigger. For clustering ($N < n$), power laws are reduced so we gain more directly calculable formulas.

Evaluating the applicability of first-order Taylor approximation, we find:

THEOREM 1. There is no difference in terms of accuracy between applying Taylor approximation at the first order to de-clustered ($N > n$) or direct length chains ($N = n$).

PROOF. See the Appendix.

This allows us to drop the case $N > n$ and independently from any Taylor development concentrate on the clustering $N = n$ and $N = 1$. The intuitive direct length representation $N = n$ has a methodological advantage that can be described as:

THEOREM 2. When applying homogenization by keeping the original number of process stages ($N = n$), the homogeneous parameters are independent of any other homogeneous parameters.

PROOF. For $N = n$, eq. (16) to (19) can be obtained respectively from eq. (7) to (10) by homogenizing them with eq. (11) to (15), or by setting a unique and constant value to heterogeneous parameters of the same kind, i.e. $\forall k = 1...n, d_k = d, e_{mk} = e_m, e_{ik} = e_i, m_k = m, i_k = i,$ etc. □

Homogenized parameters $e_m$ ($N = n$) and $e_i$ ($N = n$) thus do not depend on $d$ ($N = n$) given by eq. (12) and can be considered as fixed mean values in (16) to (19), which is an advantage of direct-length representation. A clear advantage of $N = 1$ homogenization is the linearity of eq. (16) to (19) and the resulting ease of analytically solving strategy comparisons. As a clear drawback, the homogenized parameters obtained by $N = 1$ are not independent from each other, e.g. $e_m$ ($N = 1$) and $e_i$ ($N = 1$) are non-linear functions of $d$ ($N = 1$). However, eq. (11) to (15) allow us to establish a link between parameters with different $N$, so that the numerical values of constant parameters like $e_m$ ($N = n$) and $e_i$ ($N = n$) can be exactly retrieved through non-linear transforms from $e_m$ ($N = 1$) and $e_i$ ($N = 1$). This elegant transformation can be expressed as the following:

THEOREM 3. Clustering of $N = 1$ leads to an exact numerical representation of critical parameters of the direct length chain $N = n$.

PROOF. See the Appendix.

Theorem 3 enables us to use the analytically meaningful $N = 1$ clustering, or indeed any clustering approach. Numerical values of other clustering approaches can be calculated without referring back to the original heterogeneous chain by using a double transform: first clustering of the chain using $N$ = 1, then transformation of parameters into the $N = n$ space. This has two main advantages from a methodological point of view. First, it is numerically faster than classic numerical methods, as one simply has to solve lower power equations in $N$ = 1 (numerically or theoretically) and then to rescale the solutions into the $N = n$ space. Second, whatever $N$ , homogenizing with eq. (11) to (15) keeps the boundary values of the parameters constant, e.g. if $e_m$ ( $N = n$ ) = 0 then $e_m$ ( $N$ = 1 ) = 0 and conversely.

To validate this approach, we wrote a short computational program which determines the critical parameters in the direct length representation $N = n$ for the maintenance problem outlined. With an initial set of given parameters, the program computes the unit costs according to the original heterogeneous formulas (8) to (10). Second, it homogenizes these costs and other parameters with $N = n$ for each strategy. Third, it compares the homogenized unit costs of different maintenance strategies and by means of a fixed-point iteration converges rapidly to points of unit cost equality, yielding the critical parameters.

Any maintenance strategy comparison, or indeed most complex production chain optimizations, can thus be conducted using 3 independent methods: a direct analytical study in $N = n$ (if possible), an indirect method using theorem 3 (solve in $N$ = 1, rescale to $N$ = n), and a computational program approach for a numerical study in $N = n$. In the course of our research, we applied all three methods for the comparisons of maintenance strategies: zero maintenance versus monitoring, zero maintenance versus inspection, and monitoring versus inspection. The first comparison "zero maintenance versus monitoring" is useful as we can demonstrate our methodology on a simple case and validate it with numerical simulations. The second comparison "monitoring versus inspection" applies the same methodology and yields meaningful results for the outlined production problem and will therefore be presented extensively.

## 4. Methodology Applied I: Comparison of Monitoring and Zero Maintenance

We will start our illustration of the homogenization and clustering methodology by showing a numerical example of unit costs of monitoring for different effectiveness values and the corresponding unit costs of zero maintenance in a direct length clustering scenario of $N = n$ = 50, as depicted in fig. 2. To distinguish between clustering approaches, we will in the following use superscripts to indicate the applied approach whenever needed for clarity of exposition, e.g. $e_m$ ($N = n$) will be noted $e_m^{N=n}$ etc. If monitoring effectiveness is maximal ($e_m = 1$), the defect rate does not affect unit costs of monitoring as all defects are prevented. For decreasing effectiveness, unit costs of monitoring are increasing in form of sigmoid functions (with cutoffs at 0 and 1), leveling off at a maximum of nearly double the baseline costs.

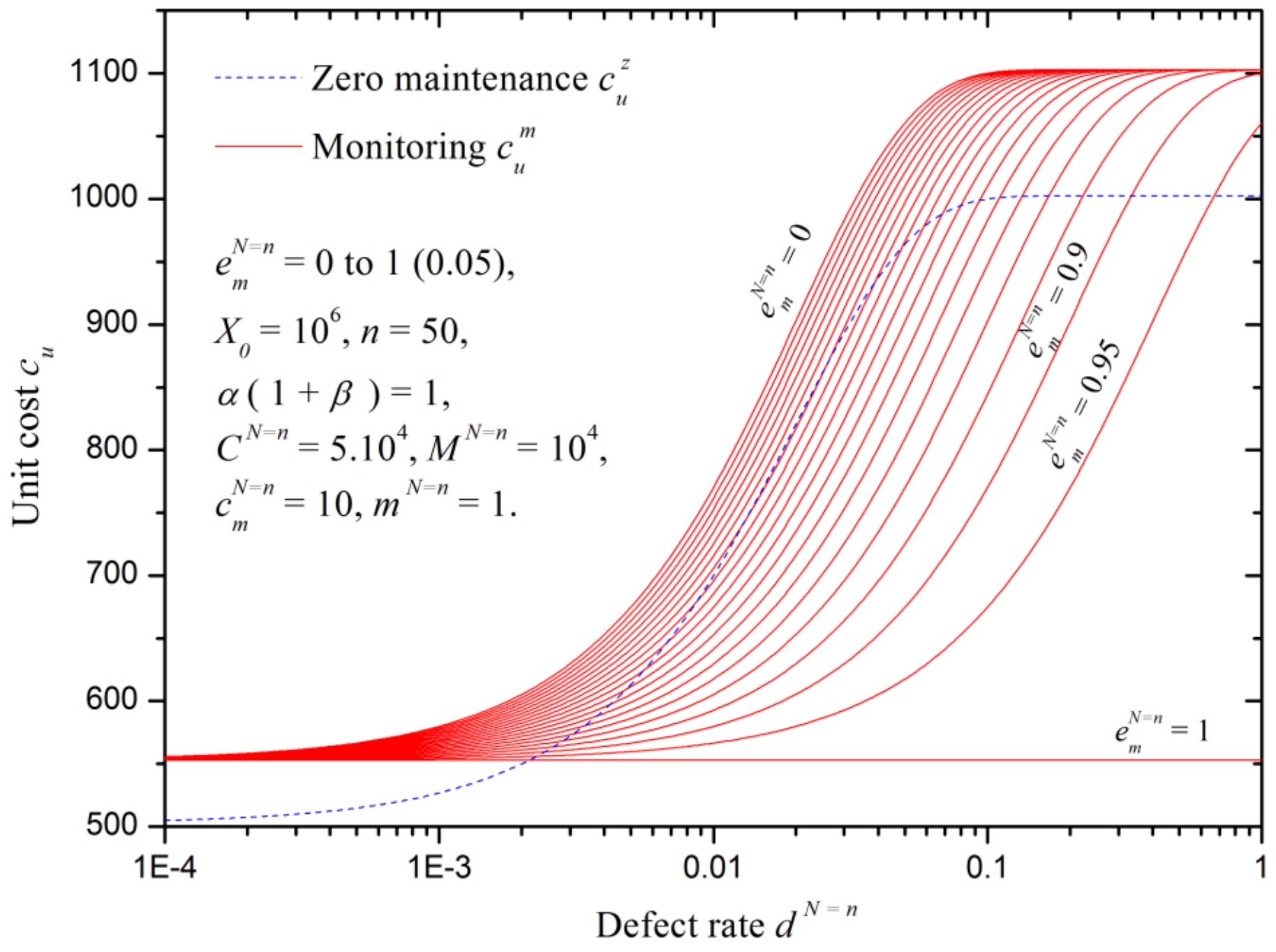

Figure 2: Unit cost development for monitoring and zero maintenance according to homogenized defect rate and different monitoring effectiveness in a direct length clustering scenario.

The unit costs of zero maintenance, as indicated by the dotted line, increase in similar sigmoid fashion with the defect rate. In effect, $c_u^z$ can be regarded as a special case of $c_u^m$ with $e_m = M = m = 0$ explaining zero maintenance competitiveness for low values of $d$. For very low $e_m$, the functions never intersect and hence zero maintenance is always cheaper than monitoring whatever $d$. Economically speaking, the maintenance costs are not covered by the economic gain of improved quality if such improvement is weak. However, there is a threshold value for $e_m$ (~ 0.35 in fig. 2) for which $c_u^z$ and $c_u^m$ are tangent at one specific $d$ (~ 0.02 on fig. 2), i.e. economic net gains of quality improvement start to be appropriated. Therefore, for $e_m$ values between this threshold and 1, the $c_u^z$ function crosses the $c_u^m$ function at two points, indicating a range of monitoring superiority between two critical defect rates.

This range is depicted as the section of the functions below zero shown in fig. 3, where the difference in unit costs $\left(c_u^m - c_u^z\right)$ is plotted against the defect rate for varying monitoring effectiveness. Visible is the increasing range of monitoring superiority for increasing monitoring effectiveness, with only a short range of very small defect rates resulting in zero maintenance superiority ($d < 2.5\times10^{-3}$).

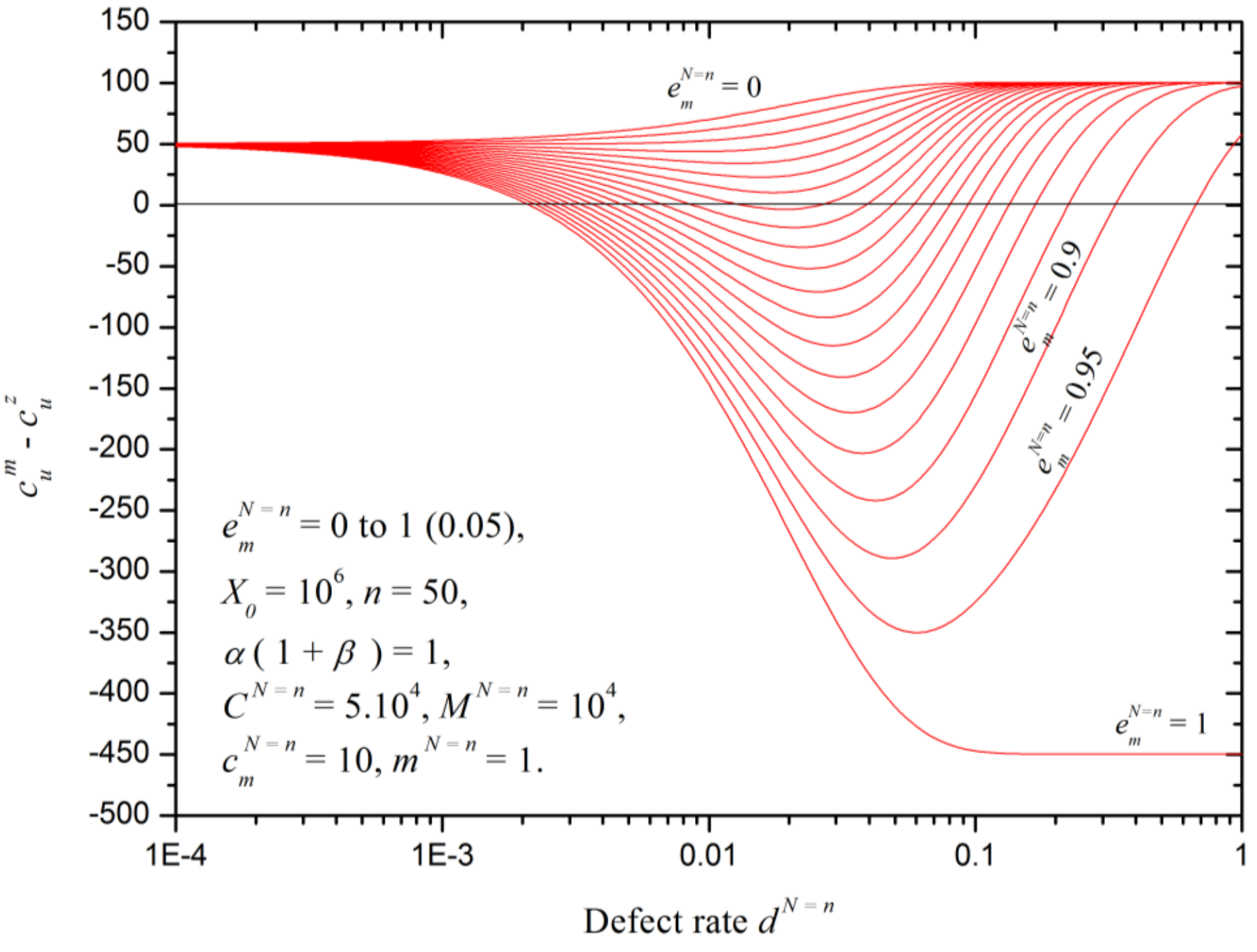

Figure 3: Difference in unit costs and range of monitoring superiority for the $N = n = 50$ example.

Considering all the points on the zero line of fig. 3, we can derive the $(e_m)_c$ function representing the critical points of $e_m$ according to $d$ where the superiority transition from zero maintenance to monitoring occurs. Shown in fig. 4, the function $(e_m)_c = f(d)$ then has a pipe shape with a minimum, which is exactly the abovementioned threshold. The domain of monitoring superiority is then the region enclosed by the curve.

The critical monitoring effectiveness is dependent on the reputation parameters $\alpha$ and $\beta$. For a strong market reaction, e.g. $\alpha(1+\beta)=4$ in fig. 4, the domain of monitoring superiority is large with a low threshold, e.g. $(e_m)_c = 0.18$ for $d = 0.012$. But decreasing market reactions shift the minimum as well as the function itself upward and to the right: for e.g. $\alpha(1+\beta)=0.2$, monitoring superiority exists only for $e_m > 0.8$ and $d > 0.015$. Practically speaking, a low market reaction to poor quality signifies quality is only of minor importance for the customer and hence there are no high transaction or goodwill costs to be considered, rendering zero maintenance advantageous.

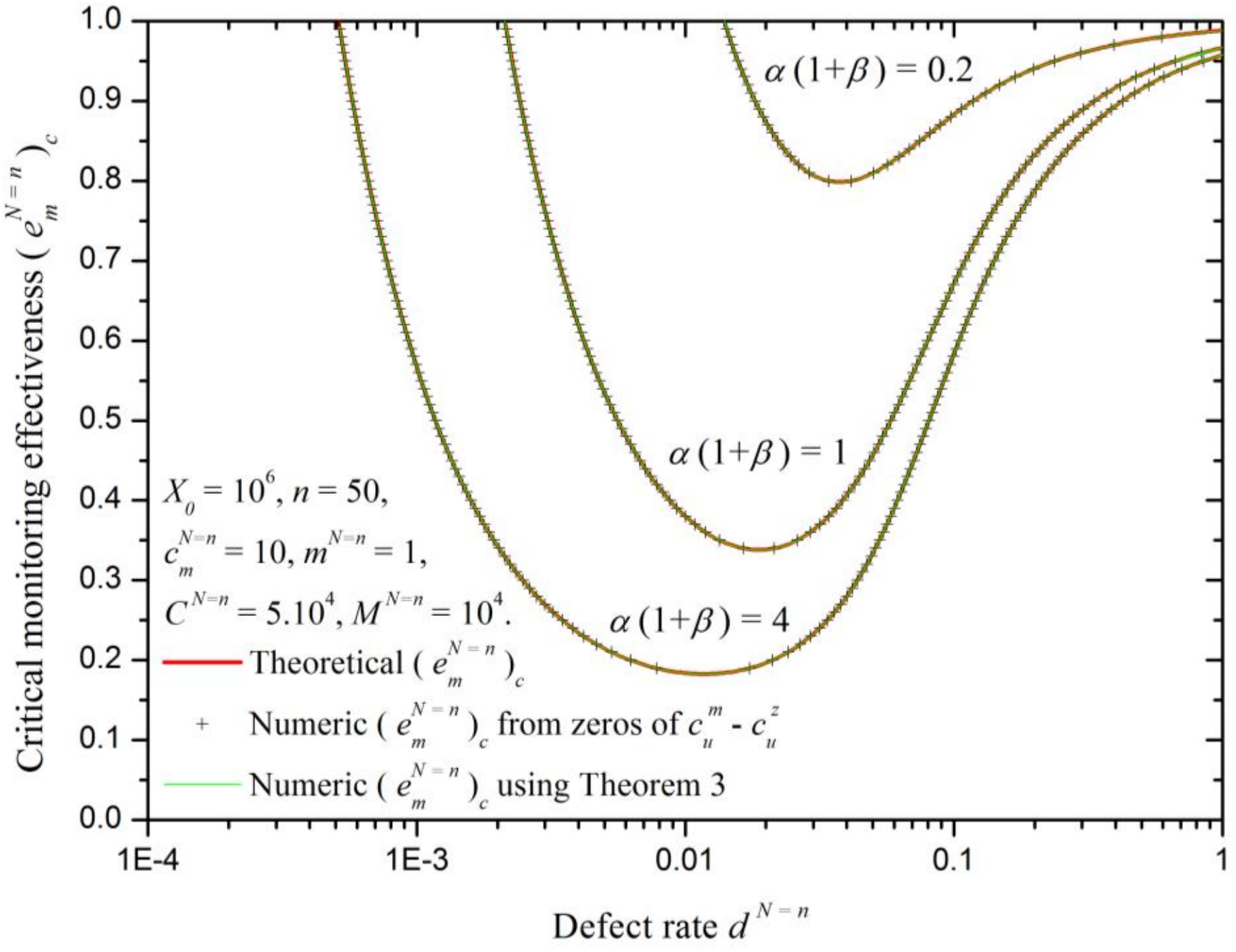

Figure 4: Theoretical and numerical derivation of the critical monitoring effectiveness for changing reputation parameters for the $N = n = 50$ example; for given reputation parameters, monitoring superiority corresponds to the domain above the curve.

Those preceding numerical results can be confirmed and extended by theoretical studies. As seen in fig. 4, the $\left(e_m^{N=n}\right)_c$ curve can be obtained by means of the three mentioned approaches to the problem – the numerical derivations from our program (search for the zeros of $c_u^m - c_u^z$), the theoretical comparisons of costs, and an indirect method using theorem 3 – which all show exactly the same results. Analytically, it is possible to directly compare eq. (18) to eq. (19) and thus to derive the critical monitoring effectiveness in $N = n$ as

(20) $$\left(e_m^{N=n}\right)_c = 1 - \frac{1 - \left(\left(\left(\frac{M^{N=n}/X_0 + m^{N=n}}{\alpha(1+\beta)} + m^{N=n} + c_m^{N=n}\left(1-d^{N=n}\right)^n\right) / \left(c_m^{N=n} + m^{N=n}\right)\right)^{1/n}}{d^{N=n}}$$

which is the theoretical curve in fig. 4. Any $e_m^{N=n} > \left(e_m^{N=n}\right)_c$ will make monitoring more advantageous than zero maintenance. The same comparison of eq. (18) to eq. (19) in $N = 1$ yields

(21) $$\left(e_m^{N=1}\right)_c = \frac{1}{1 + c_m^{N=1}/m^{N=1}} + \frac{M^{N=1}/X_0 + m^{N=1}}{\alpha(1+\beta)\left(c_m^{N=1} + m^{N=1}\right)d^{N=1}}$$

To retrieve the $N = n$ case from eq. (21), we use the following non-linear transforms

(22) $$d^{N=n} = 1 - \left(1 - d^{N=1}\right)^{1/n},\ e_m^{N=n} = 1 - \frac{1 - \left(1 - \left(1 - e_m^{N=1}\right)d^{N=1}\right)^{1/n}}{1 - \left(1 - d^{N=1}\right)^{1/n}} \text{ and } V^{N=n} = \frac{V^{N=1}}{n}$$

with $V = C$, $M$, $c_m$ or $m$, and arrive again at eq. (20). This transform is numerically much faster than any classical numerical methods, as an analytical formula like (21) is simply numerically rescaled. It is also possible to derive theoretical results, e.g. we can see from eq. (21) that $\left(e_m^{N=1}\right)_c$ decreases with $\alpha$ and $\beta$, while those parameters are not present in eq. (22), leading to the conclusion that eq. (20) must decrease with $\alpha$ and $\beta$. This method in general is adequate for complicated strategy comparisons, where the direct length formula (20) may not exist or is so complex that it is only possible to derive results from a formula such as (21). Further exploring eq. (20), (21) and (22), we can derive:

LEMMA 1. Monitoring is superior to zero maintenance in a simply connected space clearly defined by critical monitoring effectiveness and boundary values of the defect rate.

PROOF. The region where monitoring is superior to zero maintenance is defined by $e_m^{N=n} \geq \left(e_m^{N=n}\right)_c$, with $\left(e_m^{N=n}\right)_c$ given by eq. (20), and the cutoffs or boundary conditions $e_m^{N=n} \in [0,1]$ and $d^{N=n} \in [0,1]$. Eq. (20) for $\left(e_m^{N=n}\right)_c$ is a smooth regular function taking only positive values whatever $d^{N=n} \in [0,1]$ and value 1 only for $\left(d^{N=n}\right)_{\min} = 1 - \left(1 - \frac{M^{N=n} / X_0 + m^{N=n}}{\alpha\left(1+\beta\right) c_m^{N=n}}\right)^{1/n} > 0$. As for any $d^{N=n} > \left(d^{N=n}\right)_{\min}$, $\left(e_m^{N=n}\right)_c < 1$, the cutoffs of eq. (20) delimitate a single domain. □

COROLLARY 1. Whatever *n*, the monitoring superiority domain does not exist if $\alpha\left(1+\beta\right) \le \left(M / X_0 + m\right) / c_m$.

It follows from corollary 1 that in markets with little or zero external quality effect, i.e. $\alpha\left(1+\beta\right) \approx 0$, zero maintenance is always cheaper than monitoring. The fact that increasing monitoring effectiveness does not change the result is due to fact that our simplified model hypothesizes that all production is sold whatever quality level, hence neglecting multi-period sales effects. This corresponds to a typical "reputation milking" situation when a producer exploits quality uncertainty for one period, as e.g. described in Shapiro (1983). Further study leads to lemmas 2 to 4:

LEMMA 2. The monitoring superiority domain increases in $\alpha\left(1+\beta\right)$ or $c_m$, and decreases with *M* or *m*.

Lemma 2 underlines the main economic facts: the stronger the market reaction to poor quality, the higher the product's unit costs, or the lower the fixed or variable monitoring costs, the more desirable monitoring. Regarding the length of the production process, we find:

LEMMA 3. The monitoring superiority domain decreases in *n*, other things being equal.

Lemma 3 is noteworthy as it illustrates an important aspect: the longer the process chain, the more important it is to aim for high monitoring effectiveness, as a higher number of process stages *n* increases the likelihood of good units made defective in the course of production. At high uniform defect rates and long process chain, an insufficient effectiveness means that the beneficial effect of monitoring is consumed by the still-incurred waste costs and the direct costs of monitoring, making zero maintenance the more economical strategy.

LEMMA 4. $\left(e_m^{N=n}\right)_c$ as given in (20) admits one global minimal value, which is non-zero and approximately stable in *n* for *n* >> 1.

Combined with lemma 2, lemma 4 signifies that this minimal value, together with its corresponding $d^{N=n}$ and the width of the critical curve, decreases in $\alpha\left(1+\beta\right)$ or $c_m$, and increases in $M$ or $m$. Combined with lemma 3, the corresponding $d^{N=n}$ and the width of the critical curve decrease in $n$.

To summarize, for $d^{N=n} < \left(d^{N=n}\right)_{\min}$ zero maintenance is superior to monitoring, defects being so few that it is too costly to eliminate their causes, while for $d^{N=n} > \left(d^{N=n}\right)_{\min}$ monitoring is superior to zero maintenance if $e_m^{N=n} \geq \left(e_m^{N=n}\right)_c$. $\left(e_m^{N=n}\right)_c$ first decreases with $d^{N=n}$ to a minimum before increasing again, as $c_u^m$ increases for higher defect propensities while $c_u^z$ is saturated (see fig. 2). In practice, monitoring superiority is realistically achievable, e. g. $e_m^{N=n} \geq 0.5$ for $d^{N=n} \in [0.005, 0.05]$ with $\alpha\left(1+\beta\right) = 1$, and the size of its domain grows following the quantity $\alpha\left(1+\beta\right)c_m / \left(M / X_0 + m\right)$, i.e. the external quality effect adjusted by the monitoring costs.

Lemma 1 to 4 show the consistency of our model as quality maintenance makes only sense when an external effect is to be expected. In case of no market reaction, a producer sells whatever he produces – and saves the costs for maintenance. Such cases are certainly rare as most companies operate in competitive markets where customers have a choice and will switch to the competition's product if they are not compensated for experiencing bad quality. Moreover, our model allows us to give quantitative predictions of the tipping point (defined by critical parameters) between two maintenance strategies.

## 5. Methodology Applied II: Comparison of Monitoring and Inspection

The same model can be applied to the analysis of the main trade-offs between inspection and monitoring. Fig. 5 shows the unit cost development for the three strategies based on the above numerical example for direct length representation $N = n = 50$ with equal maintenance parameters like costs per stage ($M = I$, $c_m = c_i$ and $m = i$) and effectiveness $e_m = e_i = 0.8$. The functions for monitoring and zero maintenance cross for two $d$ values already seen in fig. 2.

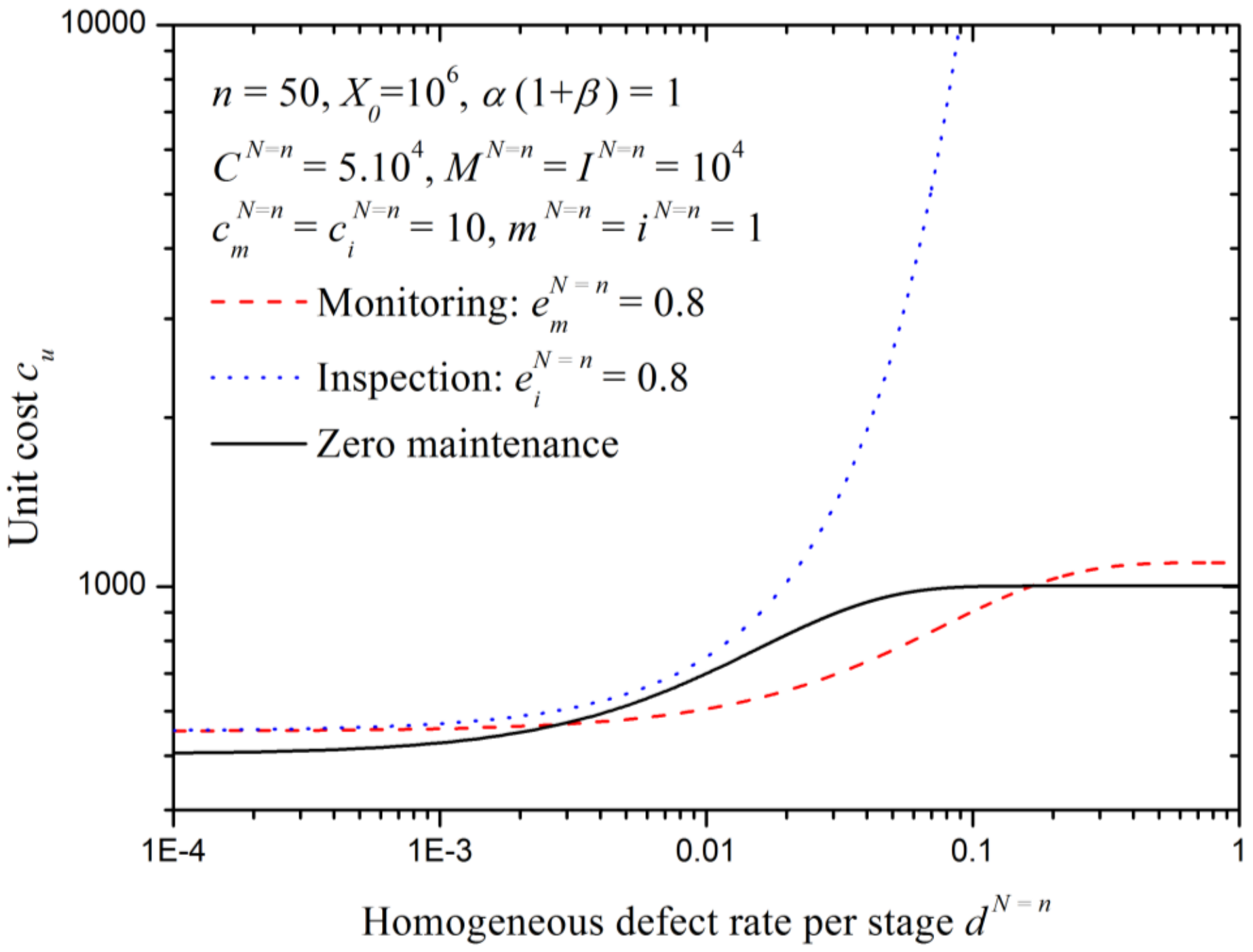

Figure 5: Unit cost development of the three pure strategies for equal maintenance effectiveness $e_m^{N=n} = e_i^{N=n} = 0.8$ based on numerical example.

In the numerical example underlying fig. 5, monitoring is superior or equally preferable to inspection irrespective of the defect rate. According to eq. (17) inspection cost are inversely proportional to $(1-e_i d)^N$ and thus exponentially increase when $e_i d \to 1$, while according to eq. (18) and (19) monitoring and zero maintenance saturate to the respective upper bounds $c_u^m = N(C+M)/X_0 + N(1+\alpha(1+\beta))(c_m+m)$ and $c_u^z = NC/X_0 + N(1+\alpha(1+\beta))c_m$. Those bounds give the values $c_u^m = 1103$ and $c_u^z = 1002.5$ in fig. 5.

Due to its exponential cost development inspection seems largely inferior in this setting, but what if we consider a larger range of potential settings? The first important aspect to assess is the potentially differing effectiveness of the two strategies, keeping the maintenance costs per stage equal. In a second step, we complement this finding by taking into account differences between the maintenance costs per stage. Finally we focus on how the reputation parameters $\alpha$ and $\beta$ affect the result.

From a methodological point of view, we use the same tools as in the previous section and focus on the critical effectiveness of monitoring. The main difference to the former section is that the direct length representation $N = n$ yields an expression of $\left(e_m^{N=n}\right)_c$ much more complex than eq. (20). To circumvent this, we will use the method based on theorem 3, which is to write all equations in $N = 1$ and then to rescale parameters to $N = n$. This will allow us to do fast computing modeling and to derive general analytical results.

Starting with the analysis of effectiveness, fig. 6 produces a detailed view of varying effectiveness values by showing the points of monitoring and inspection cost equality ( $c_u^m = c_u^i$ ) according to $e_m$, $e_i$ and $d$ for $N = n$ in a 3D plot. Numerically, these points are obtained by search of zeros of $c_u^m - c_u^i$ with a precision of $10^{-5}$ (computation time ~ 2 hours) as well as by applying theorem 3 and solving the equation in space $N = 1$ (computation time ~ 1 second). The generated surface can be regarded as the function $\left(e_m^{N=n}\right)_c$ according to $e_i^{N=n}$ and $d^{N=n}$, below which inspection is superior and above which monitoring is superior.

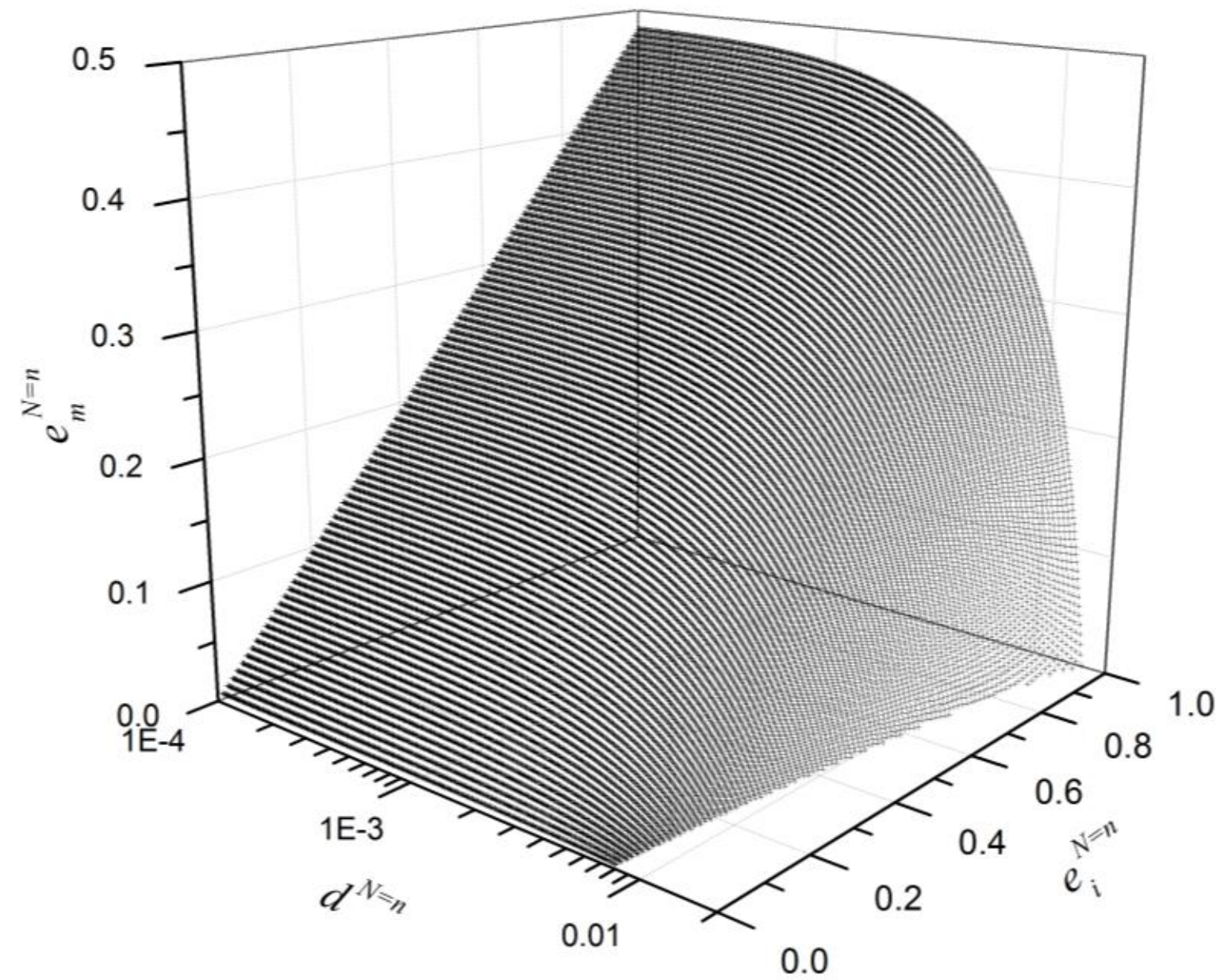

Figure 6: Points of monitoring and inspection unit cost equality according to $e_m^{N=n}$, $e_i^{N=n}$ and $d^{N=n}$, obtained by means of numerical study or theorem 3 (same parameters as in fig. 5).

Note that increasing monitoring effectiveness tips the balance in favor of monitoring at high levels of inspection effectiveness, but increasing inspection effectiveness has nearly no impact at high levels of monitoring effectiveness. This result has practical implications as it signifies that investing in monitoring is beneficial even when a company has previously run an inspection system, but not vice versa. Hence, "sunk cost" considerations should not prevent a company from gradually replacing inspection by monitoring. In production reality, quality improvement often targets only sections of the production process, and installing sensors in the recently improved process stages is clearly more beneficial than returning to an inspection scheme. Note further that the maximum of $\left(e_m^{N=n}\right)_c$ is lower than 0.5, thus signifying that, for the given parameters of fig. 6, any greater monitoring effectiveness makes monitoring the superior choice no matter how well inspection is conducted.

Analytically, using the same homogenization technique as for the derivation of eq. (21) we obtain the general expression of the critical effectiveness of monitoring for $N = 1$ as

$$\begin{aligned}\left(e_m^{N=1}\right)_c &= 1+\frac{1}{\alpha(1+\beta)d^{N=1}}+\frac{C^{N=1}+M^{N=1}}{X_0\alpha(1+\beta)\left(c_m^{N=1}+m^{N=1}\right)d^{N=1}} \\ &-\frac{\dfrac{C^{N=1}+I^{N=1}}{X_0}+c_i^{N=1}+i^{N=1}}{\alpha(1+\beta)\left(c_m^{N=1}+m^{N=1}\right)\left(1-e_i^{N=1}d^{N=1}\right)d^{N=1}}-\frac{\left(c_i^{N=1}+i^{N=1}\right)\left(1-e_i^{N=1}\right)}{\left(c_m^{N=1}+m^{N=1}\right)\left(1-e_i^{N=1}d^{N=1}\right)^2}\end{aligned} \tag{23}$$

At $\left(e_m^{N=1}\right)_c$ both maintenance strategies are equally preferable, and monitoring superiority is given whenever $e_m^{N=1} > \left(e_m^{N=1}\right)_c$. In case of equal maintenance cost parameters ($M = I$, $c_m = c_i$ and $m = i$), the study of eq. (23) shows that $\left(e_m^{N=1}\right)_c$ decreases in $d^{N=1}$. As the rescaling to $N = n$ does not change this result, we arrive at:

LEMMA 5. For equal maintenance cost parameters, whatever $N$ and $e_i \in [0;1]$, the maximum value for $\left(e_m\right)_c$ as a function of $d \in [0;1]$ is achieved for $d = 0$.

Lemma 5 allows Taylor approximation as a complementary tool since the condition $d << 1$ is fulfilled, developing the expressions of cost $c_u$ as a Taylor's series in $d$ and then neglecting the terms in $d^2$. This yields the expression of $\left(e_m\right)_c$ at small defect rates:

LEMMA 6. Whatever $N$, for $d << 1$, $\left(e_m\right)_c$ can be approximated by

(24) $$\left(e_m\right)_c = 1 - \frac{\left(c_i + i\right)\left(1 - e_i\right)}{c_m + m} - \frac{\left(\frac{C+I}{X_0} + \frac{1}{2}\left(1 + \frac{1}{n}\right)\left(c_i + i\right)\right)e_i}{\alpha\left(1+\beta\right)\left(c_m + m\right)} + \frac{\frac{M-I}{X_0} + m - i + c_m - c_i}{\alpha\left(1+\beta\right)\left(c_m + m\right)nd}$$

COROLLARY 3. The threshold values at $d$ = 0 remain constant whatever $N$.

COROLLARY 4. For $\frac{M-I}{X_0} + m - i + c_m - c_i = 0$ and $d << 1$, $\left(e_m\right)_c$ is independent of $d$, thus $\left(e_m\right)_c$ reaches an upper bond for $e_i = 1$ that represents an absolute maximum value as a function of $e_i$ and $d$:

(25) $$\text{Max}\left(e_m\right)_c = 1 - \frac{1}{\alpha\left(1+\beta\right)\left(c_m + m\right)}\left(\frac{C+I}{X_0} + \frac{1}{2}\left(1 + \frac{1}{n}\right)\left(c_i + i\right)\right)$$

To illustrate, the highest corner of the curve in fig. 6 has the value of $\text{Max}\left(e_m\right)_c \approx 0.4845$. For any small value of $d$ (i.e. below $d$ = 2.10$^{-3}$ in fig. 6), $\left(e_m\right)_c$ grows proportionally to $e_i$ with the rate $\frac{\Delta\left(e_m\right)_c}{\Delta e_i} \approx \frac{c_i + i}{c_m + m} \times \text{Max}\left(e_m\right)_c \approx 0.4845$.

COROLLARY 5. Whatever $e_i \in [0\,;1]$, for $\frac{M-I}{X_0} + m - i + c_m - c_i = 0$ the threshold value at $d$ = 0 can be calculated from eq. (24) as

(26) $$\left.\left(e_m\right)_c\right|_{d=0} = 1 - \frac{c_m + i}{c_m + m} + e_i\left(\frac{c_m + i}{c_m + m}\left(1 - \frac{1}{2}\frac{1 + 1/n}{\alpha\left(1+\beta\right)}\right) - \frac{C+I}{X_0\alpha\left(1+\beta\right)\left(c_m + m\right)}\right)$$

and can be seen as a linear function of $e_i$, i.e. $\left.\left(e_m\right)_c\right|_{d=0} = \frac{\Delta\left(e_m\right)_c}{\Delta e_i} \times e_i + 1 - \frac{c_m + i}{c_m + m}$. This threshold value delineates the critical monitoring effectiveness to be achieved in order to guarantee monitoring superiority under the condition of maintenance cost equality (whatever $d$, as $d$ = 0 gives the maximum value). Taking the numerical example of figs. 6 and 7, this equation predicts $\left.\left(e_m\right)_c\right|_{d=0} = 0.3876$ for $e_i = 0.8$ or inversely $e_i = 0.826$ if we set $\left.\left(e_m\right)_c\right|_{d=0} = 0.4$ as a given condition.

We will now focus on how these results change under different cost conditions and reputation effects. For simplicity, we assume that the condition $c_m^{N=n} = c_i^{N=n}$ is fulfilled, i.e. costs per stage before adding any maintenance cost are the same. Based on eq. (24), $\delta = \left(M^{N=n} - I^{N=n}\right) / X_0 + m^{N=n} - i^{N=n}$ is the parameter to study maintenance cost influence, and $\alpha(1+\beta)$ the one to study reputation influence. Figure 7 shows the effect of these two parameters on $\left(e_m\right)_c$ in both $N = n$ and $N = 1$ spaces.

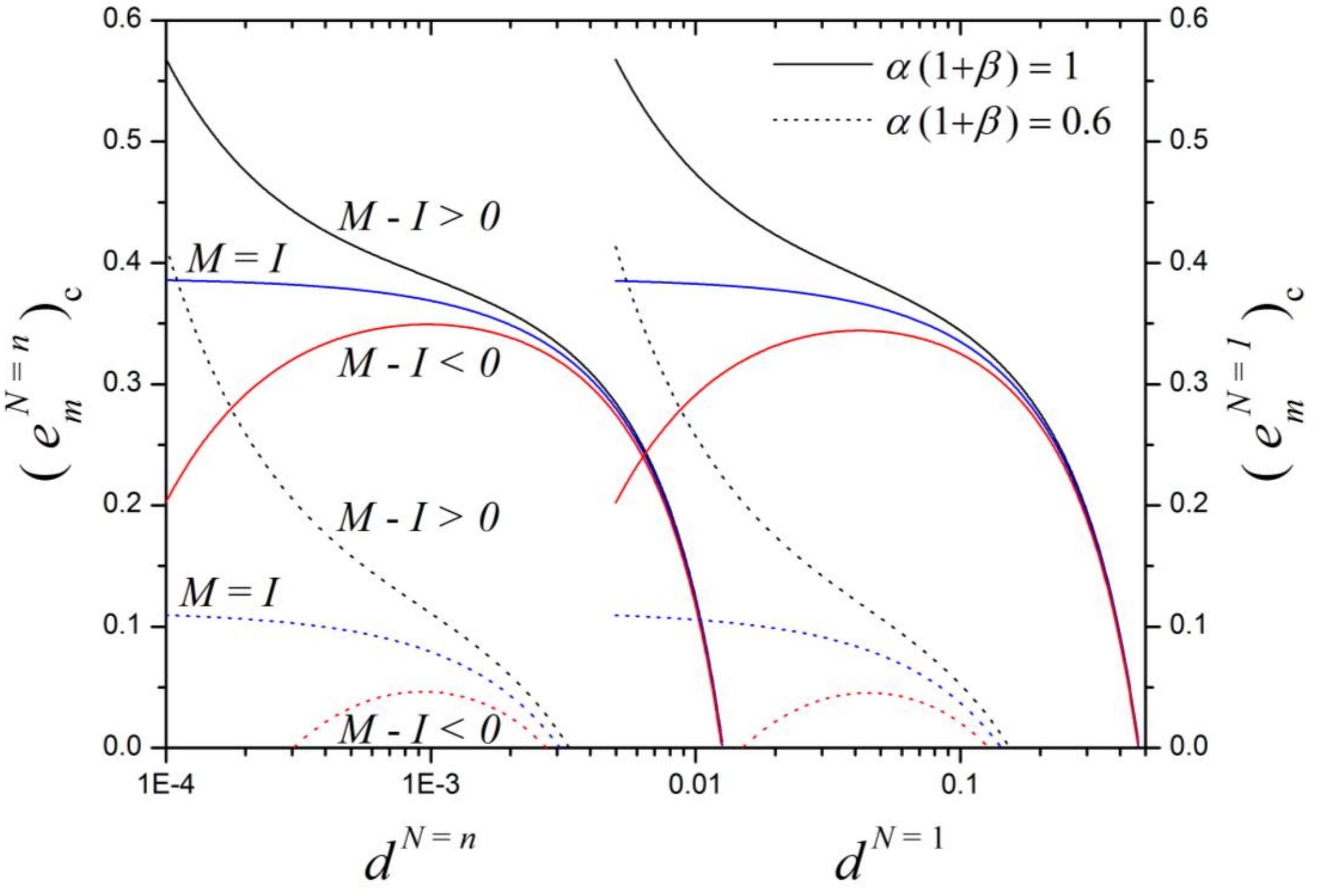

Figure 7: For $e_i^{N=n} = e_i^{N=1} = 0.8$ and *m* = *i*, $\left(e_m^{N=n}\right)_c$ as a function of $d^{N=n} \in [10^{-4}; 0.5]$ (on the left) and $\left(e_m^{N=1}\right)_c$ as a function of $d^{N=1} \in [5.10^{-3}; 0.5]$ (on the right) for 3 different values of *M* – *I* with high reputation $\alpha(1+\beta) = 1$ (plain lines) and lower reputation $\alpha(1+\beta) = 0.6$ (dotted lines).

First, the set of curves in space *N* = *n* = 50 is directly transformed into the set of curves in *N* = 1 space via theorem 3 and eq. (22), with a simulation cutoff for $\left(e_m^{N=n}\right)_c$ at $d^{N=n} = 10^{-4}$ yielding a corresponding cutoff for $\left(e_m^{N=1}\right)_c$ at $d^{N=1} = 5 \cdot 10^{-3}$. This transformation preserves exactly the shape of the curves. Second, for any given $\alpha(1+\beta)$, $\left(e_m^{N=n}\right)_c$ is higher for *M* > *I* then for *M* = *I* and lower for $M < I$ ($m$ is set equal to $i$ here). For the cost equality curve $M = I$ we retrieve the maximum value given by eq. (26) when $d$ is small, i.e. $\left(e_m\right)_c\big|_{d=0} = 0.3876$ for $\alpha\left(1+\beta\right) = 1$, and $\left(e_m\right)_c\big|_{d=0} = 0.1156$ for $\alpha\left(1+\beta\right) = 0.6$. Third, the inspection superiority domain is greatly reduced for lower reputation effect, and is of importance only for high reputation effect. We should then be able to derive a critical value of $\alpha\left(1+\beta\right)$ at which inspection superiority vanishes completely whatever $d$ (see corollary 6).

From an analytical point of view, the features of the fig. 7 can be explained by the study of eq. (23) and (24). First, we can explain the shape of the curves: for $d \ll 1$, eq. (24) gives the slope of the curves as negative if $\delta < 0$, constant if $\delta = 0$, and positive if $\delta > 0$. For a bigger $d$, we can consider eq. (23) in space $N = 1$: as parameter transforms between $N = 1$ and $N = n$ preserve the shape of the curves (see fig. 7), any conclusion in one space is valid in the other one and vice versa. The study of eq. (23) shows that for $\delta \geq 0$ the function $\left(e_m^{N=1}\right)_c$ decreases with $d^{N=1}$, while for $\delta < 0$ the $\left(e_m^{N=1}\right)_c$ curve increases with $d^{N=1}$, reaches a maximum and then decreases to 0. The same applies in the $N = n$ space.

We can also deduce from eq. (23) why $\left(e_m^{N=n}\right)_c$ grows with $\delta$ as shown in fig. 7:

LEMMA 7. Whatever $N$ and $d$, the critical monitoring effectiveness $\left(e_m\right)_c$ decreases in inspection costs $I$, $i$ and increases in monitoring costs $M$, $m$. Consequently, $\left(e_m\right)_c$ increases with $\delta$ and therefore the monitoring superiority domain decreases with $\delta$.

PROOF. Study of eq. (23) and its derivatives yield lemma 7 for $N = 1$. Transformations to other values of $N$ according to theorem 3 leave signs of derivatives unchanged. □

Lemma 7 ensures that the curves with different cost coefficients never cross in $d$, so that the maximum observed in $\left(e_m^{N=n}\right)_c$ for $\delta < 0$ cannot be higher than the value given in eq. (26), which is the upper bound of $\left(e_m\right)_c$ whatever $N$ for the case $\delta \leq 0$. On fig. 7, $\left(e_m^{N=n}\right)_c \leq 0.3876$ for $\alpha\left(1+\beta\right) = 1$ and $\left(e_m^{N=n}\right)_c \leq 0.1156$ for $\alpha\left(1+\beta\right) = 0.6$.

Finally, let us turn to the effect of reputation on unit costs. For monitoring, the reputation effect is the parameter with the greatest impact on costs, whereas the removal of defective units and hence the defect rate is the parameter with the greatest cost impact for inspection. A high $\alpha\left(1+\beta\right)$ value might thus tip the balance in favor of inspection, whereas a lower value might tip it in favor of monitoring. The behavior of $(e_m)_c$ according to $\alpha(1+\beta)$ is complex and depends on the sign of $\delta$:

LEMMA 8. For $\delta \le 0$, $(e_m)_c$ increases with $\alpha(1+\beta)$ and $e_i$ for any given *d*. For $\delta > 0$, $(e_m)_c$ increases or decreases in $\alpha(1+\beta)$ and in $e_i$ depending on *d* and costs.

PROOF. Per study of eqs. (23) and (24). □

For $\delta \le 0$, the inspection superiority domain at first increases with reputation parameters. This initial result can be explained by the number of process stages *n* concerned: in common processes the stages are sequentially linked, which means that due to the sorting-out effect of inspection, the number of units leaving the last process step and thus being sold is reduced. Hence, the absolute number of this final output being defective is smaller than the absolute number of the final output being defective when employing a monitoring system, other things being equal. This "small number" effect inflates the warranty and goodwill costs of monitoring stronger than of inspection. Higher reputation parameters simply amplify the "small number" effect. However, the total units sold for inspection get reduced with growing defect rates and the costs per unit of good product increases. Therefore, whatever $\alpha(1+\beta)$, all curves shown in fig. 7 finally decrease to reach $(e_m)_c = 0$ at $d = d_{\max}$.

COROLLARY 6. For $\delta \le 0$, there exists a minimal $\alpha(1+\beta)$ below which inspection is never superior whatever $e_i$.

To determine this minimal value, the search for $(e_m)_c \le 0$ in eq. (25) yields

(27) $$\left[\alpha(1+\beta)\right]_{\min} = \frac{1}{c_m + m}\left(\frac{C+I}{X_0} + \frac{1}{2}\left(1+\frac{1}{n}\right)(c_i + i)\right)$$

As a numerical example, the cost parameters chosen for fig. 7 give $\left[\alpha(1+\beta)\right]_{\min} = 0.51545$, a value close to $\alpha(1+\beta) = 0.6$ shown in fig. 7 where the inspection superiority domain is almost vanished.

For $\delta > 0$, the derivative $\dfrac{\partial\left(e_m^{N=1}\right)_c}{\partial\alpha\left(1+\beta\right)}$ changes signs for critical values of $e_i$ and of $\alpha\left(1+\beta\right)$. Starting with the critical value of $e_i$, we get

(28) $$\left(e_i^{N=1}\right)_c = \frac{1}{d^{N=1}}\left(1-\frac{C^{N=1}+I^{N=1}+X_0\left(c_i^{N=1}+i^{N=1}\right)}{C^{N=1}+M^{N=1}+X_0\left(c_m^{N=1}+m^{N=1}\right)}\right)$$

COROLLARY 7. For $\delta > 0$, any $e_i > (e_i)_c$ makes $\left(e_m\right)_c$ increase in $\alpha\left(1+\beta\right)$ whereas any $e_i < (e_i)_c$ has the opposite effect. At $e_i = (e_i)_c$, $\left(e_m\right)_c$ remains constant in $\alpha\left(1+\beta\right)$.

As an example, for the values $M = 2\cdot 10^4$ and $I = 10^4$ the formula (28) predicts $(e_i)_c^{N=1} = 0.353$ at $d^{N=1} = 10^{-4}$, which is the same value as for $N = n$ and can be verified numerically. For the critical value of $\alpha\left(1+\beta\right)$, considering that eq. (24) is valid for $d << 1$ whatever *N*, the derivative $\dfrac{\partial\left(e_m\right)_c}{\partial e_i}$ changes signs at

(29) $$\left[\alpha\left(1+\beta\right)\right]_c = \frac{C+I}{X_0\left(c_i+i\right)}+\frac{1}{2}\left(1+\frac{1}{n}\right)$$

COROLLARY 8. For $\delta > 0$ and $d << 1$, any $\alpha\left(1+\beta\right) > \left[\alpha\left(1+\beta\right)\right]_c$ makes $\left(e_m\right)_c$ increase in $e_i$ whereas any $\alpha\left(1+\beta\right) < \left[\alpha\left(1+\beta\right)\right]_c$ has the opposite effect. At $\alpha\left(1+\beta\right) = \left[\alpha\left(1+\beta\right)\right]_c$, $\left(e_m\right)_c$ remains constant in $e_i$.

As an example, for the values $M = 2\cdot 10^4$ and $I = 10^4$ the formula predicts $\left[\alpha\left(1+\beta\right)\right]_c = 0.5145$ at $d^{N=n} = 10^{-4}$, which is verified numerically. The logic of both corollaries 7 and 8 is that the sensitivity of $\left(e_m\right)_c$ to parameters such as $e_i$ and $\alpha\left(1+\beta\right)$ is increased when $\delta > 0$. Taking corollary 8 as an example, very low values such as $\alpha\left(1+\beta\right) < \left[\alpha\left(1+\beta\right)\right]_c$ mean there is only a negligible external quality effect, hence the unit cost impact of inspection fully takes effect. For higher reputation parameter values, this unit cost impact is overlain by the “small number” effect, which combined with the higher initial maintenance costs necessitates increasing monitoring effectiveness whenever inspection effectiveness increases.

## 6. Concluding Remarks

Illustrating the applicability of our methodological approach, sections 4 and 5 yielded critical parameters from the comparison of zero maintenance to monitoring and inspection to monitoring. As we now have two sets of results, the comparison of the deducted critical parameters reveals general results for all three strategies:

i. In a market with no reputation effect ($\alpha(1+\beta) = 0$) or in near-perfect quality condition ($d \approx 0$), zero maintenance is always the most profitable strategy.
ii. For very low values of $\alpha(1+\beta)$, zero maintenance is preferable to monitoring which in turn is preferable to inspection.
iii. Higher values for reputation parameters are associated with the need for maintenance. Monitoring and inspection are then superior to zero maintenance for most of the possible values of $d$. Inspection can outperform monitoring only if monitoring costs are relatively high, monitoring effectiveness relatively low, and in a range of small defect rates (otherwise, the unit cost effect makes inspection inferior).
iv. Zero maintenance again becomes a profitable strategy at very high defect rates (see fig. 5). Hence, monitoring is especially suited for a large middle range of defect rates (not too close to 0 and to 1).

We used the case of quality maintenance strategy comparisons to illustrate a new methodology for analyzing complex production chain problems based on the core ideas of canonical homogenization and clustering. Canonical homogenization can be generally applied to all heterogeneous multi-parameter production chain problems, which are either too difficult to analyze analytically or where statistical methods require too much computing time or produce too weak results. Using a variety of theorems, we could show that the once homogenized production functions can be clustered into four different cluster types (direct length representation, clustering, de-clustering, and the special case of total clustering $N = 1$) and that these can be transformed and mapped onto each other. Clustering leads to a non-linear transformation of homogenized parameters that substantially reduces the degree of the involved polynomial equations, i.e. it transforms a polynomial equation with a degree $n$+1 into one with a lower degree $N$+1. Hence, it reduces the number of parameters to compute, in the extreme case to only one parameter as in total clustering. Problems can be regarded in the cluster type best-suited for analysis, facilitating the finding of solutions and greatly reducing computing time, in our illustrative example by 99.98%. Although computing time might vary according to the underlying problem and the type of clustering approach, these savings give our methodology a clear advantage over traditional statistical methodologies. Another advantage is that we receive a clear result instead of a probabilistic space. Hence, it allows for running fast and meaningful simulations on complex problems.

As the business world is characterized by a multitude of repeat processes, this methodology might be useful for solving a large range of complex business problems. Most of them can be modeled quantitatively, and applying this methodology aids in finding solutions or in determining the trade-offs for making the right decisions.

We are convinced that the proposed sequence of homogenization and clustering also has value for other scientific areas as it represents a novel approach to a known class of problems. Contributing to the further refinement of the methodology, we are currently working on assessing its applicability in other fields of study. One is dynamic problem settings that change over time, which requires an extension of the heterogeneous equations. Another field is the assessment of its potential applicability in Machine Learning algorithms, which as well requires an extension of heterogeneous equations and needs to be modeled at least as a 2 dimensions system. Extending the case of quality maintenance strategies used to illustrate the methodology´s applicability, we can further study mixed strategies - i.e. a combination of both monitoring and inspection in the course of production - and extend the developed equations, such as (7) and (16). Adding additional parameters to the original heterogeneous equations, i.e. a term for stocked materials in equations (1) to (10), we can run a more detailed simulation of the economic effects of choosing a certain maintenance strategy. The same equations can also be used in further examining the criteria needed to achieve near-perfect quality levels and such enrich the theoretical fundament of Six Sigma and other quality concepts.

We hope that the methodology developed in this paper aids both researchers and practitioners in addressing chain problems in business and thereby enhance the possibilities for optimization we are all striving for.

## Appendix

PROOF OF EQUATION (1): The basic assumptions of the model are that first, all production stages (numbered 1 to *n*) are independent from each other in terms of costs, effectiveness and defective rate, and second, that maintenance activities only work in the applied stage (detection of variation in QRPs or detection of defects). Figure A1 depicts a random stage *k* of the production process and the associated parameters.

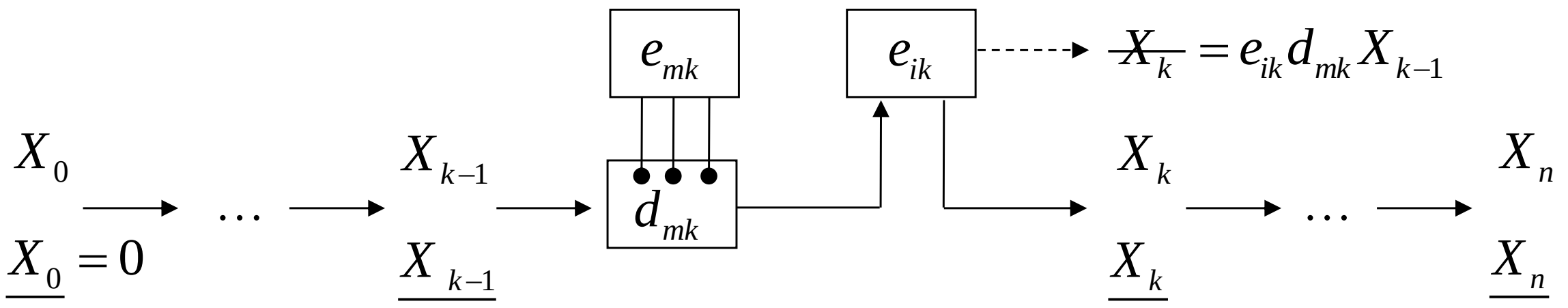

Figure A1: Stage *k* with monitoring and inspection, $d_{mk}$ being the effective defective rate after monitoring, $X_k$ the produced volume, $\underline{X}_k$ the bad volume among $X_k$, and $\cancel{X}_k$ the rejected volume.

The produced volume and its defective volume at stage *k* are calculated according to precedent stage *k*-1. The potential defective rate $d_k$ per stage *k* can first be reduced by monitoring during production to an effective defective rate $d_{mk} = (1 - e_{mk}) d_k$, and the associated defective volume is $d_{mk} X_{k-1}$. Inspection then detects defects coming from stage *k* with effectiveness $e_{ik} \in [0;1]$. Thus the detected and removed bad volume through the inspection process from that stage is $\cancel{X}_k = e_{ik} d_{mk} X_{k-1}$. The remained volume after stage *k* becomes

(A.1) $$X_k = X_{k-1} - \cancel{X}_k = X_{k-1}\left(1 - e_{ik}\left(1 - e_{mk}\right) d_k\right).$$

Following the recursive rule in eq. (A.1), the final produced volume $X_n$ after the last stage *n* can be deduced from $X_{n-1}$, which can itself be deduced from $X_{n-2}$ and so forth until the input stage $X_0$, leading finally to Eq. (1). □

PROOF OF EQUATION (2): Defective units may remain in the passed-on volume despite maintenance. At any stage *k*, its volume follows from the union of two volume sets: 1) the former defective volume $\underline{X}_{k-1}$ that passed through former stages undetected, and 2) the quantity $\left(1-e_{ik}\right)\left(1-e_{mk}\right)d_k X_{k-1}$ that is the defective volume $\left(1-e_{mk}\right)d_k X_{k-1}$ from stage *k* itself minus the quantity eliminated due to inspection $\not{X}_k = e_{ik}\left(1-e_{mk}\right)d_k X_{k-1}$. From these two sets has to be deduced the intersection set (double counted volume), i.e. $\left(1-e_{mk}\right)d_k \underline{X}_{k-1}$, which is the part of $\underline{X}_{k-1}$ that is made defective again in stage *k*. The undetected defective volume after stage *k* can thus be expressed as

(A.2) $$\underline{X}_k = \underline{X}_{k-1}\left(1-\left(1-e_{mk}\right)d_k\right) + X_{k-1}\left(1-e_{ik}\right)\left(1-e_{mk}\right)d_k .$$

Following the recursive rule in eq. (A.2), the final defective volume $\underline{X}_n$ among the total production volume $X_n$ can be expressed as in eq. (2). The proof is given by mathematical induction: Firstly, eqs. (A.2) and (2) are giving the same results for the first two terms of the series. Secondly, we show that if eq. (2) is true for a level $n \geq 2$, then it is also true for the next level $n+1$: The relation between $\underline{X}_{n+1}$, $\underline{X}_n$ and $X_n$ is provided by eq. (A.2), then replacing $\underline{X}_n$ and $X_n$ by eqs. (2) and (1) respectively yields $\underline{X}_{n+1} = X_0\left[\prod_{k=1}^{n+1}\left(1-e_{ik}\left(1-e_{mk}\right)d_k\right) - \prod_{k=1}^{n+1}\left(1-\left(1-e_{mk}\right)d_k\right)\right]$ which is eq. (2) for $n+1$. □

PROOF OF THEOREM 1: In the general homogeneous unit cost formula eq. (16), the error function between the power law terms and their first-order Taylor approximations in *d* is $O(d^2) = \left(1-e_i\left(1-e_m\right)d\right)^N - \left(1-Ne_i\left(1-e_m\right)d\right)$ with $N \in \mathbb{R}^*$. Pointing out the following property $\left(1-e_i\left(1-e_m\right)d\right)^N = \prod_{k=1}^{n}\left(1-e_i\left(1-e_{mk}\right)d_k\right) = \text{Cte}(N) \equiv \gamma$ with $\gamma \in [0,1]$, which follows directly from homogenization, the error function yields $O\left(d^2\right) = \gamma - 1 + N\left(1-\gamma^{1/N}\right)$ for $N \in \mathbb{R}^*$. As $N = 1$ only satisfies $O(d^2) = 0$ and the derivative $\dfrac{dO\left(d^2\right)}{dN} = 1 + \gamma^{1/N}\left(\dfrac{\ln\gamma}{N} - 1\right)$ is strictly positive if $\gamma < 1$ $\left[\lim_{\gamma\to 1} O(d^2) = 0\right]$, $O(d^2)$ is positive for $N > 1$ and increases with *N* up to the maximum value $\lim_{N \to +\infty} O(d^2) = -\ln\gamma + \gamma - 1$. Hence, inaccuracy due to first-order Taylor approximation with any $N > n$ is higher than with $N = n$. □

PROOF OF THEOREM 3: The aim of comparing strategies is to determine the break-even, or critical parameters, where unit costs are equal, as a function of other parameters, for example $\left(e_m^{N=n}\right)_c = f\left(d^{N=n}\right)$ in comparing monitoring to zero maintenance, $\left(e_i^{N=n}\right)_c = g\left(d^{N=n}\right)$ in comparing inspection to zero maintenance, etc. Any direct length representation $N = n$ will rarely lead to analytic expression of such functions due to the involved power law. However, any homogenized cost expression (16) to (19) written with a scaling parameter $N_1$ are also true when written with another rescaling parameter $N_2$ without referring back to the same original heterogeneous values. So, there exists a set of numerical transformations $T$ to transform homogenized parameters from $N_1$ to $N_2$: $d^{N=N_1} \xrightarrow{T} d^{N=N_2}$, $e_m^{N=N_1} \xrightarrow{T} e_m^{N=N_2}$, $e_i^{N=N_1} \xrightarrow{T} e_i^{N=N_2}$, etc. Thus, $N = 1$ clustering gives the analytical values of critical parameters, e.g. $\left(e_m^{N=1}\right)_c = f\left(d^{N=1}\right)$, which yields after applying $T$ an exact numerical representation of critical constant parameters in $N = n$, e.g. $\left(e_m^{N=n}\right)_c = f\left(d^{N=n}\right)$. □